\def\year{2019}\relax
\documentclass[letterpaper]{article} %DO NOT CHANGE THIS
\usepackage{aaai19} %Required
\usepackage{times} %Required
\usepackage{helvet} %Required
\usepackage{courier} %Required
\usepackage{url} %Required
\usepackage{graphicx} %Required
\usepackage{comment}
\usepackage{subfig}
\usepackage{csquotes} % enquotes
\usepackage{booktabs}
\usepackage{fixltx2e}
\usepackage{amsmath}

\makeatletter
\def\endthebibliography{%
 \def\@noitemerr{\@latex@warning{Empty `thebibliography' environment}}%
 \endlist
}
\makeatother

\title{Lipper: Synthesizing Thy Speech using Multi-View Lipreading}
\author{
Yaman Kumar \\ Adobe \\ykumar@adobe.com
\And Rohit Jain \\ MIDAS Lab, NSIT-Delhi\\rohitj.co@nsit.net.in
\And Khwaja Mohd. Salik\\ MIDAS Lab, NSIT-Delhi\\khwajam.co@nsit.net.in
\AND 
Rajiv Ratn Shah \\MIDAS Lab, IIIT-Delhi\\rajivratn@iiitd.ac.in
\And Yifang Yin \\NUS, Singapore \\yifang@comp.nus.edu.sg
\And Roger Zimmermann \\NUS, Singapore\\rogerz@comp.nus.edu.sg
}
 
\begin{document}
\maketitle
\begin{abstract}
Lipreading has a lot of potential applications such as in the domain of surveillance and video conferencing. Despite this, most of the work in building lipreading systems has been limited to classifying silent videos into classes representing text phrases. However, there are multiple problems associated with making lipreading a text-based classification task like its dependence on a particular language and vocabulary mapping. Thus, in this paper we propose a multi-view lipreading to audio system, namely Lipper, which models it as a regression task. The model takes silent videos as input and produces speech as the output. With multi-view silent videos, we observe an improvement over single-view speech reconstruction results. We show this by presenting an exhaustive set of experiments for speaker-dependent, out-of-vocabulary and speaker-independent settings. Further, we compare the delay values of Lipper with other speechreading systems in order to show the real-time nature of audio produced. We also perform a user study for the audios produced in order to understand the level of comprehensibility of audios produced using Lipper. 
\end{abstract}

%%%%%%%%% BODY TEXT
\section{Introduction}

\label{introduction}
Human speech is bimodal in nature, with the two modalities coming from the auditory and optical senses. An example in this regard comes from experiments by McGurk and MacDonald in their appropriately titled paper, ``Hearing lips and seeing voices'' \cite{mcgurk1976hearing}, where they show that subjects when shown mouth images speaking \emph{/ga/} but with the sound of \emph{/ba/}, perceived it as \emph{/da/}. However, the recent work in lipreading domain has decoupled the visual signals from the auditory ones. Instead, most of the lipreading projects treat this problem as a classification task where they consider speech videos from a restricted vocabulary of a particular language. Then, models are trained to classify those videos into a fixed number of classes made up of that limited vocabulary. However, scaling that approach to multiple languages and a complete vocabulary is a difficult task. In addition, humans do not always speak valid statements. Meaningless, gibberish or non-language and vocabulary-conformant speech (for instance, a human speaker making animal sounds) cannot be modeled using a restrictive approach like a classification model. Thus, with this in mind, we present a model, namely Lipper, which given a silent video consisting of lip-movements \textit{reconstructs} the speech of the speaker. It does this by modeling lipreading as a regression rather than a classification task.

\subsection{Distinction between Speech-reconstruction, Speech-reading, and Speech-recognition systems}
\label{speech_reconstruction_vs_others}
%Lipper is a \textbf{speech reconstruction} model. People can easily get confused between speech reading, reconstruction and recognition systems. The only commonality amongst all of these systems, pertaining to our work, is that lipreading has been shown to be effective in all the three tasks. Speech recognition systems simply perform the task of identifying whether a speech is of a particular person or not, and therefore, are not related to this work. 
Lipper is a \emph{speech reconstruction} model. People can easily get confused between speech reconstruction, reading and recognition systems. The only commonality amongst all of these systems, pertaining to our work, is that lipreading has been shown to be effective in all the three tasks. 

Taking the case of speech recognition systems, they perform the task of identifying the speaker of a speech, and therefore, are not related to this work. However, lipreading based speech-reading and reconstruction systems do share some common features. Thus, with this in mind, in explaining this work, we focus on speech-reading and speech-reconstruction systems \emph{only}. 

While on one hand, speech-reading systems involve identifying \textbf{what} a person says (using text as the identification metric), on the other hand, the objective of speech-reconstruction systems is to \textbf{generate} the \emph{speech} of a person (using audio as the output generated). However, generation of the speech using reconstruction systems may \emph{not} involve identifying \emph{what} a person says. For instance, speech can be generated even for illegible tokens (such as any random permutation of characters), but due to the vocabulary dependence of speech-reading models, it becomes difficult to identify these illegible tokens. %However, for the generation of speech, reconstruction models \emph{need not} know \emph{what} the speech is, they can just remake the speaker's speech (e.g, 
%For instance, consider a television, it plays the sounds of different shows from various languages without knowing which language is the show about. % This fact also manifests itself in the form of Lipper being vocabulary and language independent model. Speech reading has been based on a highly constrained lexicon of words which, then subsequently translates to a restriction on the total number of classes (\emph{i.e}, phrases, words and sentences) that are considered for the classification task. However, speech reconstruction system remakes audio for a subject based on lip movement of the subject. 
 Consider another example where this might be useful, multilingual people often speak in code-switched languages (for e.g., in India, a code-switched version of \textit{``I was in London the last month''}, can be, ``\emph{Pichle mahine}, I was in London''). For identifying that kind of speech, a speechreading system has to consider all the possible combinations of words in both the languages and then identify them. This quickly becomes un-scalable as the number of languages and the size of each vocabulary increase. % but a speech-reconstruction system can reconstruct the speech without taking the language into consideration since the lip-movements in both the languages are expected to be the same.

The reason for the language and vocabulary independence of speech reconstruction models like Lipper is that for the generation of a sound, lip, nose, throat and other \textit{movements} are required, but not the vocabulary (or language) per se. Thus, following this reasoning, one may directly translate and map lip movements to speech without referencing word or sentence mappings. %Thus, we may directly associate lip movements to sound without taking into consideration language anywhere in the pipeline. 
This is what Lipper does. Additionally, unlike other models which have to wait for the complete sentence to get over before they can start speechreading, Lipper can begin to produce audio as soon as it detects some lip-movements, thus becoming a near real-time system. Since it is not contingent on any particular language or vocabulary mapping, thus it is a \emph{language and vocabulary independent} model.

In this paper, we present an exhaustive set of experiments on Lipper for speech-reconstruction. Lipper can take multiple views into account and then lipread them to produce sound. First, in Section Speaker-Dependent Results, we present thorough experiments for exploring the quality of speech-reconstruction on all possible combinations of all the views. Second, in Section Speaker-Independent Results, using the best-view combination, we explore the results of Lipper for speaker-independent settings. Third, in Section Speaker Dependent OOV Results, we note the results of Lipper for out-of-vocabulary (OOV) phrases. Fourth, in Section Comparison of Delays, we compare the delays for speech-reconstruction and speech-reading systems thus demonstrating the difference of time taken between them. Fifth, in Section User-Study, we do a user study for the speech-comprehensibility of the sound generated. Sixth, in Section Demonstration of Reconstructed Audios, we provide links to some of the videos which show speaker-dependent and speaker-independent results produced using Lipper. Seventh, though the audios produced by Lipper might be noisy in some cases, but they do capture and contain the content of the speech of a speaker. Thus, in order to show that, we present text-classification results in Section Text-Prediction Model on the encoded audios produced. In the end, before concluding the paper, we present the future research directions in lipreading domain.

\begin{figure*}[h!]
\centering
\includegraphics[scale = 0.35]{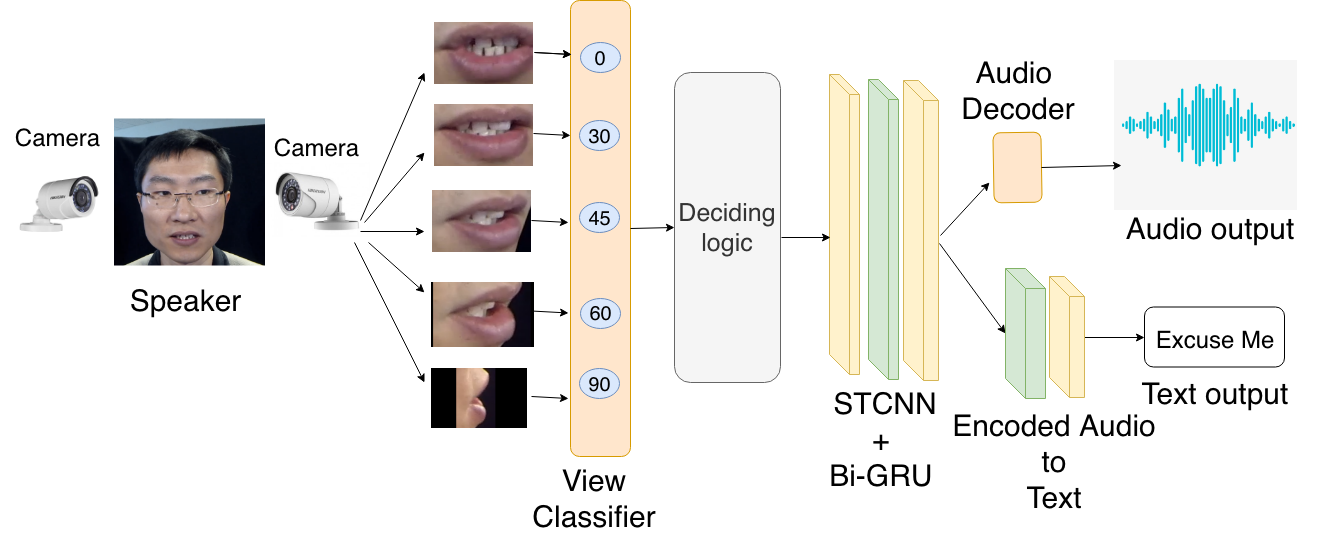}
\caption{End-to-end diagram for Lipper}
\label{fig:lipper_end_to_end}
\end{figure*}

\section{Related Work}
\label{related_work}

%Literature on lipreading spans centuries of work with the earliest documentation reported in the 17th century \cite{bulwer_1648}. The very first automated lipreading system was developed in the 1940s \cite{petajan1984automatic}. The lipreading systems have graduated several levels in terms of feature extraction, lip-tracking and accuracies obtained since then. 

Despite much research in lipreading domain, it is still seen as a classification task in which, given some silent videos, a model has to classify those videos into a limited and fixed size of lexicon \cite{lucey2006lipreading,ngiam2011multimodal,lee2016multi,zimmermann2016visual,assael2016lipnet,chung2016lip,petridis2017end,chung2017lip,shah2017multimodal}. There have also been a few works on speech-reconstruction as well \cite{cornu2015reconstructing,ephrat2017vid2speech,kumar2018harnessing,kumar2018mylipper}. However, the problem of view and pose-variation has been dealt by a very few lipreading systems \cite{zhou2014review}. 

Most of the authors have worked on frontal-view lipreading only. Lipreading on just frontal view is a major problem since a speaker cannot be expected to always face the camera while speaking. In the speechreading domain, there have been a few works which have worked on some views other than the frontal view \cite{lucey2006lipreading,kumar2007profile,lan2012view,lee2016multi,saitoh2016concatenated} but dealing with pose-variation is still a challenge. In addition, the problem is compounded by the absence of multi-view datasets. A very limited number of datasets exist for fostering research in multi-view lipreading. One of such datasets is Oulu-VS2 \cite{anina2015ouluvs2} which provides five different views of speakers shot concurrently. On this dataset, combination of multiple poses was tried for speechreading by \cite{petridis2017end} and for speech-reconstruction by \cite{kumar2018harnessing,kumar2018mylipper}. Given visual feeds from multiple cameras, the authors showed that combining multiple views would result in better accuracy in speechreading \cite{lucey2006lipreading,zimmermann2016visual,lee2016multi,petridis2017end} and better speech quality \cite{kumar2018harnessing,kumar2018mylipper} for speech-reconstruction.

Another task in lipreading domain is dealing with pose-variation. Usually, different models are made for dealing with different poses \cite{lucey2006lipreading,kumar2007profile,lan2012view,lee2016multi,saitoh2016concatenated}. The other approach for dealing with pose-variation is to extract pose-invariant features and then use them in speech-reading \cite{lucey2006lipreading,lucey2008continuous,lan2012view,estellers2011multipose}. However, the chief limitation of these systems is their low accuracies which prevents their usage. 
There have been very few works on speech reconstruction using single view visual feed \cite{ephrat2017vid2speech,cornu2015reconstructing}. However, as noted earlier, neither were the systems tested on pose variation, nor speaker-independent settings or on multiple views. We show Lipper's performance on pose-invariant multi-view speech-reconstruction.

\section{Lipper: Design and Development}
\label{Lipper_design}

\begin{figure*}[htbp!]
\centering
\includegraphics[scale = 0.27]{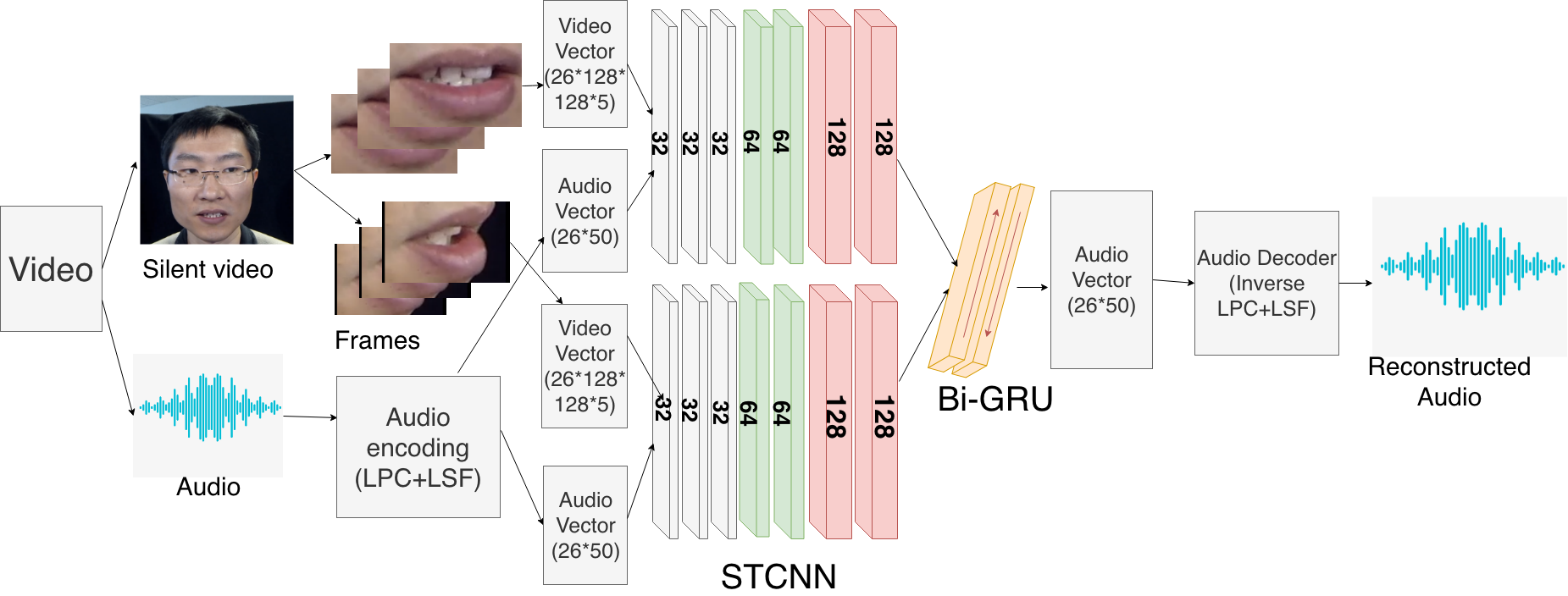}
\caption{STCNN and BiGRU based architecture used for speech reading and reconstruction }
\label{fig:stcnn_gru_model}
\end{figure*}

In this section, we describe the architecture of Lipper (as shown in Figure \ref{fig:lipper_end_to_end}). %For the construction of Lipper, the authors have experimented with multiple designs within the same architecture. Results are shown for the architectural design that gave the best results for the reconstructed audio - STCNN+BiGRU (as shown in Figure \ref{stcnn_gru_model}) based model. 
Primarily, Lipper is composed of a view classifier followed by a STCNN+BiGRU (a combination of Spatio-Temporal Convolutional Neural Network and Bidirectional Gated Recurrent Units) network (as shown in Figure \ref{fig:stcnn_gru_model}). As shown in the diagram, the view-classifier takes input from multiple cameras, and then, maps the speaker view (can be from frontal view, i.e., 0$^{\circ}$ to profile view, i.e., 90$^{\circ}$) to the nearest pose from the pose-set \{0$^{\circ}$, 30$^{\circ}$, 45$^{\circ}$, 60$^{\circ}$ and 90$^{\circ}$\}. Once this has been mapped, based on the view mapping provided by the classifier, the decision logic decides on two issues:
\begin{enumerate}
  \item Which view combinations to utilize. This is based on the experiments shown in Section Speaker-Dependent Results. The decision logic may or may not decide to utilize all the available data.
  \item Based on the above decision, it chooses the appropriate speech-reconstruction model which takes the multi-view visual input feed and generates speech.
\end{enumerate}

The system generates two types of outputs: audio and associated text. The audio is generated after decoding the encoded audio produced using the neural network. The text is generated by taking encoded audio as input in another neural network which performs classification of the encoded audio into predefined categories.

\subsection{Classifier Model}
\label{classifier_model_section}

\begin{table}
\centering
\begin{tabular}{|c|c|c|c|c|c|}
\hline
  True $^{\circ}$ / Pred. $^{\circ}$ & 0$^{\circ}$ & 30$^{\circ}$ & 45$^{\circ}$ & 60$^{\circ}$ & 90$^{\circ}$ \\ \hline
  0$^{\circ}$ & 1546 & 13 & 1 & 0 & 0 \\
  30$^{\circ}$ & 76 & 1399 & 78 & 7 & 0 \\
  45$^{\circ}$ & 3 & 19 & 1500 & 38 & 0 \\
  60$^{\circ}$ & 0 & 0 & 20 & 1538 & 2 \\
  90$^{\circ}$ & 0 & 0 & 0 & 2 & 1558 \\ 
  \hline
\end{tabular}
\caption{Confusion matrix for the view classifier}
\label{table:confusion_matrix_view_classifier}
\end{table}

\begin{figure}
\centering
\includegraphics[scale=0.45]{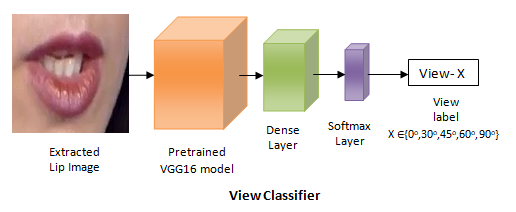}
\caption{View-classifier model for Lipper. It classifies lip region images into five categories from the set \{0$^{\circ}$, 30$^{\circ}$, 45$^{\circ}$, 60$^{\circ}$ and 90$^{\circ}$\} }
\label{fig_view_classifier}
\end{figure}

The classifier model uses transfer learning to classify lip-poses. It consists of a VGG-16 model \cite{simonyan2014very} pretrained on ImageNet images followed by one dense layer with 1024 units and then by one softmax layer with five units. The VGG-16 model helps in extracting the visual features from the lip region images. Since this is a multi-class classification problem, we use the cross-entropy loss to train the system. A visual representation of the classification model is given in Figure \ref{fig_view_classifier}. For training the model, we used lip-region images of size 224x224. While training, we use the batch size as 100 and then we train the system for 30 epochs with Adam optimization. The confusion matrix of the classifier is given in the Table \ref{table:confusion_matrix_view_classifier}. For training and testing, we used a uniform class distribution with equal number of samples from each of the classes. The overall accuracy as calculated from Table \ref{table:confusion_matrix_view_classifier} is 96.7\%.

\subsection{Speech-Reconstruction Model}
STCNN+BiGRU network helps Lipper to deal with video based data. STCNN layers extract the visual features while BiGRU layers help it to take care of temporal nature of the data. For the audio features, we use Linear Predictive Coding (LPC) \cite{fant2012acoustic} for representing audio. LPC order was found out to be optimal at 24. Line Spectrum Pairs (LSPs) \cite{itakura1975line} can represent LPC coefficients in a quantization robust manner. 
The network takes input images of lip-region of size 128x128 and produces the output as LPC+LSP encoded audio. The network consists of 7 layers of STCNN (of size 32, 32, 64, 64, 128 and 128 respectively) followed by 2 layers of BiGRU (of size 64 and 32 respectively). This is finally followed by the output layer of size 50 which produces the encoded audio.

We use 60 epochs for training and 20 epochs for fine-tuning the network. We first sample audio at a sampling rate of 20,000 and then encode it using LPC order of 24. This is then fed to the network along with the image sequence in timesteps of 5 during the training time.

For different experiments Lipper's training happens in different formats:
\begin{enumerate}
  \item For speaker dependent experiments, Lipper was trained on all possible combination of views (5 views available in total, thus $\binom{5}{1}$+$\binom{5}{2}$+$\binom{5}{3}$+$\binom{5}{4}$+$\binom{5}{5}$ possible combinations) for each individual speaker such that out of three phrases for a given speaker in the database \cite{anina2015ouluvs2}, it was trained on two and one was kept for testing the system. The results corresponding to that are given in Section Speaker-Dependent Results.
  
  \item For speaker independent settings, the best view combination as obtained from the experiments above was taken. The model then was trained on all but two speakers for all their phrases. Then, it was tested on the two speakers left out for the speech generated. The results corresponding to that are given in Section Speaker-Independent Results. 
  
  \item For out-of-vocabulary settings, Lipper was trained individually for all the speakers. The training strategy followed was such that it was trained ten times and in each iteration, one of the ten phrases was left out of the training data, and was included in the test data. Thus, in each iteration, Lipper was tested on a phrase which it had never seen. The results corresponding to that are given in Section Speaker-Dependent OOV Results.
  
\end{enumerate}

\subsection{Text-Prediction Model}

The text-prediction model takes encoded audio of the reconstructed speech as input and classifies those encoded audios into text classes. For collecting the input for this model, we used a pre-trained speech-reconstruction STCNN+Bi-GRU network and then obtained the output of \textit{all} the silent videos present in the database.

The network has four fully connected layers (with sizes as 1000, 500, 100 and 10) with dropout (0.5) after the layers with sizes 500 and 100. The loss function used was cross-entropy loss and the optimizer as Adam. The network was trained with batch size as ten and number of epochs as twenty.

For making a text predicting model, we tried two different train-test data configurations:
\begin{enumerate}
  \item In the first configuration, we randomly divided all the encoded audios of all the speakers into train, test and validation data with the ratio as (70, 10 and 20) respectively. 
  \item In the second configuration, we took \textit{all} the encoded audios of 70\% of the speakers as the training data, and divided the rest 30\% speakers' audios into 10\% and 20\% to be used as validation data and test data respectively.
\end{enumerate}
The results for both these configuration compared with those of \cite{petridis2017end} (which is the state-of-the-art speachreading model on OuluVS2) are presented in Table~\ref{table:text_prediction_model}. We present the best accuracy reported in the paper by \cite{petridis2017end} for comparison. As shown in the table, data configuration-1 performs much better than even the best-performing model of \cite{petridis2017end}. We believe this is due to the train-test configuration of Lipper itself. Randomly classifying the video into one of the ten classes present in the database would have led to 10\% accuracy. Thus, even considering the second configuration, the accuracy of the text-prediction model is non-trivial. This implies that the audios produced by Lipper, although might be noisy in some cases, capture the content of the speech of a speaker.
\begin{table}
\centering
\resizebox{.95\columnwidth}{!}{%
\begin{tabular}{|c|c|c|c|}
\hline
  & Config-1 & Config-2 & \cite{petridis2017end} \\ \hline
  \textbf{Accuracy} & 97.0 & 78.5 & 95.6 \\
  \hline
\end{tabular}
}
\caption{Accuracy of the two configurations of the text-prediction model trained compared with the best results as reported in \cite{petridis2017end}}
\label{table:text_prediction_model}
\end{table}

\section{Evaluation}
\label{evaluation}

\subsection{Database}
\label{database}

For training and testing Lipper, we use \textit{all the speakers} of OuluVS2 database \cite{anina2015ouluvs2} for speech-reconstruction purposes. OuluVS2 is a multi-view audio-visual dataset with 53 speakers of various ethnicities like European, Indian, Chinese and American. These 53 speakers speak 10 phrases with five cameras recording them simultaneously from five different angles. The angles considered are \{0$^{\circ}$, 30$^{\circ}$, 45$^{\circ}$, 60$^{\circ}$ and 90$^{\circ}$\}. The speakers speak at different pace and also stop in between while speaking. A listing of all the phrases is given in Table \ref{table_phrases}. Thus, this dataset serves well for the task at hand. This database has been used in other similar studies as well \cite{ong2011learning,zhou2011towards,pei2013unsupervised,rekik2014new,rekik2015unified,petridis2017end,kumar2018harnessing,kumar2018mylipper}.

\begin{table}[h!]
\centering
\caption{List of phrases uttered by speakers in OuluVS2}
\label{table_phrases}
\begin{tabular}{ll} \hline
\textbf{S.No.} & \textbf{Phrases} \\ \hline
\textbf{1.}  & Excuse me \\
\textbf{2.}  & Goodbye\\
\textbf{3.}  & Hello\\
\textbf{4.}  & How are you \\
\textbf{5.}  & Nice to meet you \\ \hline
\end{tabular}
\begin{tabular}{ll} \hline
\textbf{S.No.} & \textbf{Phrases} \\ \hline
\textbf{6.}  & See you \\
\textbf{7.}  & I am sorry\\
\textbf{8.}  & Thank you\\
\textbf{9.}  & Have a good time\\
\textbf{10.}  & You are welcome\\ \hline
\end{tabular}
\end{table}

\subsection{Evaluation Metric}

We choose Perceptual Evaluation of Speech Quality (PESQ) \cite{rix2001perceptual} as the evaluation metric to judge the quality of the sound. This metric has been used by other speech-reconstruction works as well \cite{ephrat2017vid2speech,ephrat2017improved,kumar2018harnessing,kumar2018mylipper}. PESQ is a ITU-T recommended standard for evaluating speech quality of 3.2 kHz codecs \cite{recommendation2001perceptual}. For comparing two audios, PESQ first level aligns them, then after passing them through filters time aligns them. It then passes the audios through auditory transform and finally extracts two distortion parameters which denote the difference between the transform of the two signals. Finally, these signals are aggregated in frequency and time, and are mapped to a MOS (mean opinion score). The range of PESQ varies from -0.5 to 4.5, where speech quality increases with increasing score.

%\subsection{Training-Testing Configuration}
%In the speaker-\textbf{dependent} experiments, we used two-thirds of the videos of \emph{a particular speaker} for training, and the remaining one-third for testing. Thus the training-testing ratio was two to one for these. For the speaker-\textbf{independent} experiments, we trained the model on all the videos for the fifty speakers and tested the system on all the videos for the rest two.

%\subsection{Training}
%\label{training_Lipper}
%This subsection presents the various training experiments conducted while building Lipper. 
%Here, we present training of Lipper on single view and multi-view visual feeds in addition to presenting the training and testing of Lipper on Out of Vocabulary (OOV) phrases.

\section{Results}
This section presents the results obtained for speaker-dependent, speaker-independent, out-of-vocabulary settings for Lipper. Additionally, it also reports the results for delay measurements on Lipper.%and the accuracy obtained on getting text out of the audio generated using Lipper. 

\subsection{Speaker-Dependent Results}
\label{sec:speaker_dependent_results}

\subsubsection{Speaker Dependent Single-View Results}
\label{speaker-dependent-single-view-results}

\begin{table*}
\centering
\caption{Mean readings for \textbf{single-view} PESQ scores for Lipper and \cite{ephrat2017vid2speech} }% on the OuluVS2 database for the first 25 speakers}
\label{1_view_results}
\begin{tabular}{llllll}
\hline
\textbf{Views} & 0 degree & 30 degree & 45 degree & 60 degree  & 90 degree  \\ \hline
\textbf{Lipper} & \textbf{2.002} & 1.750 & 1.642 & 1.744 & 1.804 \\ \hline
\textbf{\cite{ephrat2017vid2speech}} & \textbf{1.72} & 1.57 & 1.48 & 1.46 & 1.52 \\ \hline
\end{tabular}
\end{table*}

The results for Lipper when trained and tested on single-view visual feeds are shown in Table \ref{1_view_results}. We have compared our results with \cite{ephrat2017vid2speech} which train a similar speech-reconstruction network and tested it on single-view visual feed. As can be seen from the table, our results are better for all the views than \cite{ephrat2017vid2speech} but the general trend of PESQ scores is similar in both the systems. In both the models, frontal view outperforms all other views and obtains a PESQ score of 2.002 and 1.72 respectively. %The second view, i.e, 30$^{\circ}$ also closely follows frontal view and gets a PESQ score of 1.908. This is also in accordance with the results obtained by some other works \cite{lan2012view}.% In Figures \ref{orig} and \ref{1view}, the waveform and spectrogram for original audio for "Excuse Me" has been compared with what was obtained using the system trained for frontal view configuration.

\subsubsection{Speaker Dependent Multi-View Results}
\label{speaker-dependent-multi-view-results}

Table \ref{2_view_results} presents the PESQ scores for all the possible two-view combinations. As can be seen, the combination of 0$^{\circ}$ and 45$^{\circ}$ outperforms all the other possible combinations. Closely following it is the combination of 0$^{\circ}$ and 30$^{\circ}$. It should also be noted that the PESQ scores, in general, for all the possible views have been benefited after a combination with some other view. For instance, 30$^{\circ}$ in combination with 0$^{\circ}$, experiences a gain in PESQ by over 6\%. Thus with respect to placement of two cameras, in regards to obtaining best quality of audios, one should place the cameras at an angle of 45$^{\circ}$ between them. This, as shown by the table, would produce an audio which would carry the maximum quality.% In Figure \ref{2view}, the waveform and spectrogram for the audio for "Excuse Me" is presented. This was obtained using the 0$^{\circ}$ and 30$^{\circ}$ views combination.

\begin{table}[]
\centering
\caption{Mean readings for \textbf{double-view} PESQ scores}% on the OuluVS2 database for the first 25 speakers}
\label{2_view_results}
\begin{tabular}{ll} \hline
\textbf{View Union}  & \textbf{PESQ} \\ \hline  
0$^{\circ}$+30$^{\circ}$ & 2.125\\
0$^{\circ}$+45$^{\circ}$ & \textbf{2.130}\\
0$^{\circ}$+60$^{\circ}$ & 1.952\\
0$^{\circ}$+90$^{\circ}$ & 1.982\\ 
30$^{\circ}$+45$^{\circ}$ & 1.991\\ \hline
\end{tabular}
\begin{tabular}{ll}\hline
\textbf{View Union}  & \textbf{PESQ} \\ \hline  
30$^{\circ}$+60$^{\circ}$ & 1.842\\
30$^{\circ}$+90$^{\circ}$ & 2.021\\
45$^{\circ}$+60$^{\circ}$ & 1.960\\
45$^{\circ}$+90$^{\circ}$ & 1.930\\
60$^{\circ}$+90$^{\circ}$ & 1.920\\ \hline
\end{tabular}
\end{table}

Table \ref{3_view_results} presents the mean PESQ scores for all possible three view combinations. The views combination 0$^{\circ}$, 45$^{\circ}$ and 60$^{\circ}$ outperforms all the combinations and presents the best results obtained till now. This is a stupendous increase of over 32\% for both 45$^{\circ}$ and 60$^{\circ}$ when considering their single view PESQ scores only. Moreover, even while considering the combination of 45$^{\circ}$ and 60$^{\circ}$, their association with 0$^{\circ}$ leads to a non-trivial increase of more than 18\%. Although, it can be noted that not all possible 3 view combinations experience a gain over their 2 view counterparts. This might be because of less training data available due to which a larger network could not be trained appropriately. %In Figure \ref{3view}, the waveform and spectrogram for the audio for "Excuse Me" is presented. This was obtained using the 0$^{\circ}$, 45$^{\circ}$ and 60$^{\circ}$ views combination.

\begin{table}
\centering
\caption{Mean readings for \textbf{triple-view} PESQ scores }%on the OuluVS2 database for the first 25 speakers}
\label{3_view_results}

\begin{tabular}{ll}
\hline
\textbf{View Union}  & \textbf{PESQ} \\ \hline 
0$^{\circ}$+30$^{\circ}$+45$^{\circ}$ & 1.975\\
0$^{\circ}$+30$^{\circ}$+60$^{\circ}$ & 2.112\\
0$^{\circ}$+30$^{\circ}$+90$^{\circ}$ & 2.005\\
0$^{\circ}$+45$^{\circ}$+60$^{\circ}$ & \textbf{2.315}\\
0$^{\circ}$+45$^{\circ}$+90$^{\circ}$ & 1.814\\ \hline
\end{tabular}
\begin{tabular}{ll}
\hline
\textbf{View Union}  & \textbf{PESQ} \\ \hline 
0$^{\circ}$+60$^{\circ}$+90$^{\circ}$ & 1.987\\
30$^{\circ}$+45$^{\circ}$+60$^{\circ}$ & 1.931\\
30$^{\circ}$+45$^{\circ}$+90$^{\circ}$ & 1.903\\
30$^{\circ}$+60$^{\circ}$+90$^{\circ}$ & 1.965\\
45$^{\circ}$+60$^{\circ}$+90$^{\circ}$ & 1.838\\ \hline
\end{tabular}
\end{table}

Results for all possible 4-view combinations are shown in Table~\ref{4_view_results}. In most cases, there is not a major increase in PESQ scores from the three-view combinations or in some cases, even a decrease in the scores is observed. As has been stated above, the reason for this decline in performance can be due to non-availability of adequate data for training the larger network. However, some view combinations have better scores than their individual view counterparts.% In Figure \ref{4view}, the waveform and spectrogram for the audio for "Excuse Me" is presented. This was obtained using the 0$^{\circ}$, 45$^{\circ}$, 60$^{\circ}$ and 90$^{\circ}$ views combination.

Table \ref{5_view_results} shows the result obtained on all-view combinations. As can be seen, although the network at this stage outperforms all of its single view counterparts but does not perform as good as the best possible three view combination of 0$^{\circ}$, 45$^{\circ}$ and 60$^{\circ}$. We have not compared 2-views, 3-views, 4-views and 5-views combinations with models by other authors since as mentioned in the Section Related Work, there were no models previously who have worked on combinations of multiple views.

\begin{table}
\centering
\caption{Mean readings for \textbf{quadruple-view} PESQ scores}% on the OuluVS2 database for the first 25 speakers}
\label{4_view_results}
\begin{tabular}{ll}
\hline
\textbf{View Union}  & \textbf{PESQ} \\ \hline 
0$^{\circ}$+30$^{\circ}$+45$^{\circ}$+60$^{\circ}$ & 2.11\\
0$^{\circ}$+30$^{\circ}$+45$^{\circ}$+90$^{\circ}$ & 1.916\\
0$^{\circ}$+45$^{\circ}$+60$^{\circ}$+90$^{\circ}$ & \textbf{2.147}\\
0$^{\circ}$+30$^{\circ}$+60$^{\circ}$+90$^{\circ}$ & 2.071\\
30$^{\circ}$+45$^{\circ}$+60$^{\circ}$+90$^{\circ}$ & 1.948\\ \hline
\end{tabular}
\end{table}

\begin{table}
\centering
\caption{Mean reading for \textbf{all-view} PESQ scores}
\label{5_view_results}
\begin{tabular}{ll}
\hline
\textbf{Combination of views}  & \textbf{Mean PESQ scores} \\ \hline 
0$^{\circ}$+30$^{\circ}$+45$^{\circ}$+60$^{\circ}$+90$^{\circ}$ & \textbf{2.086}\\ \hline
\end{tabular}
\end{table}

\subsection{Speaker Dependent OOV Results}
\label{OOV_results}

\begin{table}
\centering
\caption{Mean readings for single view PESQ on the OuluVS2 database for out of vocabulary (OOV) phrases }
\label{table:oov_results}
\begin{tabular}{ll}
\hline
\textbf{Phrases} & \textbf{Mean PESQ scores} \\
\hline
Excuse Me  & 1.79 \\
Goodbye   & 1.66 \\
Hello  & 1.82 \\
How are you & 1.84 \\
Nice to meet you & 1.57 \\
See you   & 1.64 \\
I am sorry  & 1.68 \\
Thank you  & 1.55 \\
Have a good time & 1.46 \\
You are welcome & 1.60 \\
\hline
\end{tabular}
\end{table}

\begin{comment}
\begin{figure}[h!]
\centering
\includegraphics[scale = 0.3]{OOV_2_views.png}
\caption{Comparison of OOV results on eight audios for male (Speaker 1) and female (Speaker 10) speakers (randomly chosen) obtained using the overall best model (0$^{\circ}$, 45$^{\circ}$ and 60$^{\circ}$ combination) }
\label{oov_2_views}
\end{figure}
\end{comment}

One of the major strong points for any speech reconstruction system is their ability to reconstruct speech on phrases which were not present in the training set. For a system which treats lipreading as a classification task, this is not possible since essentially, these systems have to mark any video with the limited classes that they consider. However, human language (or, in general, sounds made by humans) presents a very wide vocabulary and cannot be modeled easily with speechreading models. 

In the Table \ref{table:oov_results}, we present the PESQ scores using the overall best model (0$^{\circ}$, 45$^{\circ}$ and 60$^{\circ}$ combination), for each of the phrases from the Table \ref{table_phrases} considered as out-of-vocabulary in different iterations. The OOV PESQ scores though lesser than their speaker-dependent counterparts are not inconsequential.

\subsection{Speaker Independent Results}
\label{speaker_independent_results}
This section presents the speaker independent results obtained using Lipper. We do not evaluate the speaker independent results on every combination of multiple views. We choose those combinations which prove to be the best in speaker dependent settings. The results for the male and female speakers (Speakers 38 and 39, respectively), are presented in the Table \ref{speaker_independent_results_table}. It can be observed that the results for speaker dependent models are significantly better than the speaker independent one. We believe this is so since Lipper does not only depend on lip movements of individual speakers but also their voices. The lip-movements of the speakers although are different, but carry the commonality of movement while speaking the same words. However, since Lipper depends not only on lip-movements but also voices of the speakers and since the voice of each speaker is different, thus the model does not perform well on the PESQ score evaluation for unknown speakers. Thus, in speaker independent settings, Lipper is not able to learn the person-specific voice features which are crucial for getting high scores using PESQ as the evaluation metric. This can explain the noticeable difference between the speaker dependent and independent results obtained using Lipper. Additionally, it can also be noted that results for male speaker are better than the female one, we believe this is so since the number of male speakers in the dataset are in a majority thus forming a bias for Lipper in favour of male voice generation.

\subsection{Comparison of Delays}
\label{delay_comparison}
One of the major advantages of Lipper is it being a near real-time system. In this section, we compare the end-to-end delay in getting speech from Lipper and a similar work of that of speechreading by \cite{petridis2017end} thus confirming the validity of Lipper being a small delay speech-reconstruction system. The comparison of delay values is reported in Table \ref{delay_results}. It is to be noted that the delay values for speechreading work of \cite{petridis2017end} depend on the phrase spoken, longer the phrase higher is the delay. %For both the systems, we calculate the delay between getting the first 

\begin{table}[]
\centering
\caption{Readings on the best-view combination PESQ scores for speaker-independent settings}% on the OuluVS2 database for the first 25 speakers}
\label{speaker_independent_results_table}
\begin{tabular}{lll} \hline
\textbf{View Union}  & \textbf{Male} & \textbf{Female}\\ \hline  
0$^{\circ}$ & 1.90 & 1.76\\
0$^{\circ}$+45$^{\circ}$ & 2.03 & 1.85\\
0$^{\circ}$+45$^{\circ}$ + 60$^{\circ}$ & 1.94 & 1.86\\
0$^{\circ}$+45$^{\circ}$ + 60$^{\circ}$ + 90$^{\circ}$ & 1.91 & 1.82\\
0$^{\circ}$+30$^{\circ}$+ 45$^{\circ}$ + 60$^{\circ}$ + 90$^{\circ}$ & 1.91 & 1.83\\
\hline \end{tabular}
\end{table}

\begin{table}
\centering
\caption{Mean readings for Delay values (in secs) on Lipper and \cite{petridis2017end} }
\label{delay_results}
\begin{tabular}{lll}
\hline
\textbf{Video } & \textbf{Lipper(s)} & \textbf{Petridis et al.(s)} \\
\hline
Excuse Me  & 0.169 & 1.26\\
Goodbye   & 0.169 & 0.94\\
Hello  & 0.169 & 0.98\\
How are you & 0.169 & 1.24\\
Nice to meet you & 0.169 & 1.4\\
See you & 0.169 & 1.34\\
I am sorry & 0.169 & 1.44\\
Thank you & 0.169 & 1.09\\
Have a good time & 0.169 & 1.61\\
You are welcome & 0.169 & 1.95\\
\hline
\end{tabular}
\end{table}

\subsection{User-Study}
\begin{table}
\centering
\caption{User studies for the reconstructed audios}
\label{table:user_study_results}
\begin{tabular}{lll}
\hline
\textbf{Study } & \textbf{Accuracy} & \textbf{Variance} \\
\hline
\textbf{Audio-only}  & 80.25 & 2.72\\
\textbf{Audio-visual}   & 81.25 & 1.97\\
\hline
\end{tabular}
\end{table}

Although PESQ is a standard numeric measure which can give an idea of the virtue of a speech and can give a reference metric for comparison but from our experiments, it was observed that the measurements produced by PESQ were not always perfect. Even for some noisy audios, the PESQ scores were high and the vice-versa was also found out to be true. Due to these observations, a user study was done to understand the intelligibilty of the audios. Two types of user studies were done for doing this analysis:
\begin{enumerate}
  \item In the first study, we gave reconstructed speech of 25 speakers to 10 different annotators who were given 4 options consisting of 1 true and 3 random options (amongst the 10 classes). Each annotator was asked to listen to the audios produced by the model as many times as he likes and then choose the best option amongst the 4 classes given. 
  
  \item In the second study, in accordance with a real video-conference environment, we showed the speaker's videos to the annotators with the reconstructed audio playing along with the video. Then, the annotators were asked to choose among 10 phrase classes for the audio-video sequence that they just listened to. 
\end{enumerate}

The results for both the user studies are given in Table \ref{table:user_study_results}. In the audio-only study, a dice throw could get 25\% of the annotations right but the annotation accuracy turned out to be 80.25 with annotator accuracy variance value as 2.72\%. In the audio-visual study, although a dice-throw would have led 10\% annotations correct, Lipper achieved an annotation accuracy of 81.25 with inter-annotator accuracy variance as 1.97\%. 

\subsection{Demonstration of Reconstructed Audios}
\label{demo_reconstructed}
Just numeric results cannot do justice to reconstructed \emph{speech} output. Thus, we have made a video listing consisting of all the reconstructed speech outputs as part of a \emph{Youtube} channel. The readers are encouraged to view the video playlist at \url{https://www.youtube.com/playlist?list=PL9rvax0EIUA4LNaXSeVX5Kt6gu2IBBnsg}. This playlist contains videos consisting of speech reconstructed using speaker dependent model, speaker independent model, OOV phrases and videos of some non-dataset speakers who speak multiple languages (Hindi, English and Chinese). The non-dataset speakers and the languages they speak show the language and vocabulary independence of the model. %We have uploaded these videos as part of \emph{Anonymous} channel, which cannot be linked anyhow with the authors' identity.
For them, we train speaker-dependent models, and first carefully get their lip-region videos and then reconstruct the speech using the model generated. Please use headphones to be able to listen to the reconstructed speech better. It is worth noting that in the demonstration\footnote{We obtain the reconstructed videos with the best 3-view (0$^{\circ}$, 45$^{\circ}$ and 60$^{\circ}$ combination). The audios are played three times so that readers can easily understand them.}, the audio is in complete \emph{sync} with the video. In addition, the speaker's voice is comprehensible and can be understood.

\section{Conclusion and Future Work}
\label{conclusion}

\subsection{Future Research Directions}
As explained in this paper, not much research has happened in speech reconstruction domain. Thus, there are a lot of areas where speech-reconstruction system can be improved. 

As shown in the Section Demonstration of Reconstructed Audios, the audios are robotic in nature. One of the main reasons for this is that voice is generated not just using mouth, but also using nose, throat and tongue. Since Lipper takes only lip-region into account, thus the voice generated cannot have emotion, prosody or modulation. Therefore, speech reconstruction systems have to work to make the audios more real-life.

Currently, the system only works in controlled environment where speakers are not moving much and are looking into the camera at a stable angle. However, in the real world scenario, this cannot be the case. The speakers will turn and twist and their poses would vary dramatically, thus, going forward speech-reconstruction has to take that into account.

In this paper, although speaker-independent settings were explored, but as was seen, the system does not work very well on them. This is a major problem for the deployment of speech-reconstruction systems in their use-case scenarios.

\subsection{Conclusion}
In this paper, the authors proposed a real-time, language and vocabulary independent, multi-view accounting and speaker-independent speech reconstruction system, namely Lipper, which utilizes multi-view visual feeds to generate the speech of a speaker. Lipper extracts features directly from the pixels of the multi-view videos. It then learns those spatial features jointly along with temporal features to finally reconstruct speech of a user. The proposed system showed significant intelligibility for the audios constructed. The best combination of views was found to be the combination of 0$^{\circ}$, 45$^{\circ}$ and 60$^{\circ}$. This combination produced a significant gain over other possible views and their combinations. We also showed the experiments of out-of-vocabulary phrases for speech reconstruction and the delay between getting speech for speachreading and speech-reconstruction systems.

\section*{Acknowledgement}
This research was supported in part by the National Natural Science Foundation
of China under Grant no.~61472266 and by the National University of Singapore
(Suzhou) Research Institute, 377 Lin Quan Street, Suzhou Industrial Park, Jiang
Su, People's Republic of China, 215123. 

MIDAS lab gratefully acknowledges the support of NVIDIA Corporation with the donation of a Titan Xp GPU used for this research.

\fontsize{9.0pt}{10.0pt}
\selectfont
\bibliographystyle{aaai}
\bibliography{bibliography}
\end{document}